\documentclass[12pt]{article}
\usepackage{epsfig}

\def\beq{\begin{equation}}
\def\eeq#1{\label{#1}\end{equation}}
\def\eeqn{\end{equation}}

\def\beqa{\begin{eqnarray}}
\def\eeqa#1{\label{#1}\end{eqnarray}}
\def\eeqan{\end{eqnarray}}

\let\bar=\overbar

\def\Dslash{\not{\hbox{\kern-4pt $D$}}}
\def\dslash{\not{\hbox{\kern-2pt $\del$}}}

\def\msb{{\bar{\ssstyle M \kern -1pt S}}}

\def\Title#1{\begin{center} {\Large {\bf #1} } \end{center}}

\begin{document}

\Title{Crystalline Color Superconductivity in Dense Matter}

\bigskip\bigskip


\begin{raggedright}

{\it Deog Ki Hong\index{Hong, D. K.}\\
Department of Physics\\
Pusan National University\\
Pusan 609-735, KOREA}
\bigskip\bigskip
\end{raggedright}

\section{Introduction}

In this talk I will discuss a recently proposed color superconducting
phase of asymmetric quark matter where the up and down quark have
different chemical potential, being in chemical equilibrium with
electrons. Using Schwinger-Dyson equations derived
from an effective theory of low-energy quasiparticles, we examine
both the case of an effective four Fermi interaction (appropriate
for intermediate densities, such as those found in a neutron star)
and Landau-damped one gluon exchange (appropriate for asymptotic
density)~\cite{Hong:2001cv}.
We then briefly discuss patching together regions of plane-wave
(LOFF) condensation.

One of the most intriguing problems in QCD is to understand how
matter behaves at extreme densities, densities much higher than
the nuclear  density, $\rho_0=0.16 ~{\rm fm^{-3}}$, as
expected in the core of compact stars like neutron stars or in the
relativistic heavy ion collision (RHIC). According to QCD, which is
now firmly believed to be the theory of strong interaction, the
interaction among quarks becomes weaker and weaker as they get closer
and closer. Therefore,
when nucleons are closely packed, the wavefunctions
of quarks inside nucleons will overlap with each other at high
nucleon density and hadronic matter will turn into quark matter.
The physical properties of quark matter are questions we
wish to answer.

In general, weakly interacting fermion matter will be a Fermi liquid,
forming a Fermi surface.
However, according the Cooper theorem,
the IR fixed point of the system of weakly interacting fermions
will be very different from the Fermi liquid,
if there exists an attraction between fermions
with opposite momentum, even for an arbitrarily weak attraction.
For quark matter, quarks attract not only holes but also quarks themselves
if scattering occurs in the color anti-triplet channel.
This can be seen by calculating the Coulomb potential due to one-gluon
exchange interaction, which is valid at high density;
\begin{equation}
V(r)=-i\int {{\rm d^3q}\over (2\pi)^3}
~T^AD_{00}^{AB}(\vec q,q_0=0)T^Be^{i\vec q\cdot\vec x},
\end{equation}
where $T^A$'s are $SU(3)_c$ generators and $D_{\mu\nu}^{AB}$ is
the gluon propagator. In perturbative QCD,
$D_{00}^{AB}=i\delta^{AB}/{\vec q}^2$ and we see that the sign of the Coulomb
potential depends on the initial (or final) color states $i,j$ (or $k,l$)
of quarks:
\begin{eqnarray}
T^A_{ik}T^A_{jl}=-{2\over 3}\delta_{[ij]}\delta_{[kl]}
+{1\over 3}\delta_{(ij)}\delta_{(kl)}.
\end{eqnarray}
Of course, the Coulomb potential by one-gluon exchange interaction would
not apply at intermediate density where the strong coupling constant is
no longer small. However, it is quite reasonable to assume that
the color exchange interation is attractive even at the intermediate
density for the color anti-triplet
channel since it reduces the color Coulomb energy among quarks.
Since attraction occurs in the
color anti-triplet channel for the diquark scattering
and in the color singlet channel for the quark-hole scattering,
the candidates for the condensates are therefore either diquark condensates
in color anti-triplet channel or quark-hole condensates in color singlet
channel, given as
\begin{eqnarray}
\left<\psi_i(\vec p)\psi_j(-\vec p)\right>\ne0{\quad}{\rm or}{\quad}
\left<\bar\psi_i(-\vec p)\psi_j(\vec p)\right>\ne0.
\end{eqnarray}
Note that the quark-hole condensate (or density wave)
carries a momentum $2\vec p$.
If it were translationally invariant condensate, it would involve
antiquarks and thus energetically not prefered to form such a condensate.
Since the diquark condensate has zero total momentum, the whole Fermi
surface can contribute to the diquark scattering amplitude, while
the phase space of quark-hole scattering is only a small fraction of the
Fermi surface due to the momentum conservation. Indeed, it is shown that
the diquark condensate is energetically more preferred to
density waves, though the color exchange attration is weaker, due
to the big difference in the phase
space~\cite{Shuster:2000tn}.

Depending on the density, the diquark condensate in quark matter
takes two different forms. At the intermediate density,
where the strange quark is too heavy to participate in Cooper pairing,
the condensate takes
\begin{eqnarray}
\left< \psi_{Li}^{a}(\vec p)\psi_{Lj}^{b}(-\vec p) \right>
=-\left<\psi_{Ri}^{a}(\vec p)\psi_{Rj}^{b}(-\vec p)\right>
=\epsilon_{ab}\epsilon^{ij3}\Delta,
\end{eqnarray}
where $a,b$ are the flavor indices, running from up and down quarks.
In this phase, called the two-flavor superconducting (2SC) phase,
the condensate is flavor singlet but color anti-triplet.
The condensate does not break any
flavor symmetry except the $U(1)$ baryon number, while breaking
the $SU(3)$ color gauge group down to a $SU(2)$ subgroup.
Since the Cooper-paring quarks should have
equal and opposite momenta,  the minimal energy needed to
Cooper-pair a strange quark with up or down quarks is, if we neglect
the interaction energy,
\begin{equation}
\delta E=\sqrt{p_F^2+m_s^2}-p_F\simeq {m_s^2\over 2\mu},
\end{equation}
where $p_F\simeq \mu$ is the Fermi momentum of light quarks, almost
equal to the quark chemical potential.
We therefore see that the energy we gain by pairing the strange quark
with light quarks, which is called Cooper-pair gap, $\Delta_0$,
has to be bigger than the minimal energy $\delta E$ we have to provide
for strange quarks to participate in Cooper-pairing.

On the other hand, at high density, where
$\mu>m_s^2/(2\Delta_0)$, it is energetically
prefered for the strange quarks to form Cooper pairs with
light quarks. At such high density, it is shown that the condensate
takes a so-called color-flavor locking (CFL)
form~\cite{Alford:1998mk,Evans:1999at,Hong:1999ru}
\begin{eqnarray}
\left< \psi_{Li}^{a}(\vec p)\psi_{Lj}^{b}(-\vec p) \right>
=-\left<\psi_{Ri}^{a}(\vec p)\psi_{Rj}^{b}(-\vec p)\right>
=k_1\delta_i^a\delta_j^b+k_2\delta_j^a\delta_i^b,
\nonumber
\label{cfl}
\end{eqnarray}
which breaks not only the color symmetry but also the chiral symmetry,
leaving only the diagonal subgroup $SU(3)_V$ unbroken.

A natural place to look for the signature of color superconductivity
is a dense object like compact stars. However, since the stars are
electrically neutral, the chemical potentials of $u,d,s$ quarks are
not equal if electrons are present. One therefore must
take into account the effect of flavor asymmetry in Cooper pairing among
different flavors.
At weak interaction equilibrium,
$u+e^-\leftrightarrow d\,(s)+\nu$,
the up and down quarks have different chemical potentials,
$\mu_d-\mu_u(\equiv2\delta\mu)=\mu_e$, if
the electron chemical potential is nonzero, $\mu_e\ne0$.
When the Fermi surface mismatch becomes large enough,
the input energy to make the momentum of
pairing quarks equal and opposite exceeds the Cooper gap
and the BCS pairing breaks down. The critical chemical potential,
at which the BCS pairing breaks, is shown to be
$\delta\mu=\Delta_0/\sqrt{2}$~\cite{lo}. However,
as shown by Larkin and Ovchinnikov~\cite{lo}, and also
by Fulde and Ferrell~\cite{ff}, even at $\delta\mu>\Delta_0/\sqrt{2}$,
diquark condensate is possible, if we allow the diquark pair to carry
a momentum~\cite{Alford:2001ze}, ${2\vec q=\vec p_u+\vec p_d}$,
$\left< \psi_{u}^{i}(\vec p_u)\psi_{d}^{j}(\vec p_d) \right>
=\epsilon^{ij3}\Delta(\vec q)$.

For such pairing, the (effective) four-Fermi interaction
is not marginal and thus does not lead to Landau pole or
dynamical mass unless the interaction is strong
enough, which is a characteristic feature
in dimensions higher than (1+1)~\cite{critical}.
As we will see later, LOFF pairing indeed occurs
in dense QCD with light flavors when
the couplings are bigger than critical values for both high and
intermediate density.

\section{Intermediate Density}
At intermediate densities, where the effective QCD coupling is
large and long-range interactions are likely to be screened, we
take the Lagrangian to be
\begin{equation}
{\cal L}=\bar\psi\left(i\!\!\!\not\partial +
\mu\gamma^0\right)\psi+{G\over2}\left(\bar\psi\psi\right)^2.
\end{equation}
Following the high density effective theory~\cite{Hong:2000tn},
we introduce
the Fermi-velocity ($\vec v_F$) dependent field, defined as
\begin{equation}
\psi(\vec v_F,x)\equiv e^{-i\mu\vec v_F\cdot \vec x}\psi(x),
\end{equation}
to rewrite the Lagrangian
in terms of particles near the Fermi surface:
\begin{equation}
{\cal L}=\sum_{\vec v_F}\psi^{\dagger}(\vec v_F,x)iV\cdot \partial
\psi(\vec v_F,x)+\sum_{\vec v_F^u,\vec v_F^d}{G\over2}
\bar\psi\psi(\vec v_F^u,x)\bar\psi\psi(\vec v_F^d,x)+\cdots
\end{equation}
where $V^{\mu}=(1,\vec v_F)$, and the ellipse denotes other
four-Fermi operators involving different Fermi velocities and
higher order terms in $1/\mu$ expansion. Note that the velocity
dependent field carries the residual momentum $l^{\mu}$, if the
quark carries momentum $p^{\mu}=(l_0,\mu\vec v_F+{\vec l}\,)$.

Introducing auxiliary fields, $\sigma(x)$, we rewrite the
interaction Lagrangian as
\begin{equation}
{\cal L}_{4F}=\sigma(\vec q,x)\psi(\vec v_F^d,x)\psi(\vec v_F^u,x)-
{1\over 2G}\sigma^2,
\end{equation}
where $2\vec q=\mu_u\vec v_F^u+\mu_d\vec v_F^d$ is a fixed vector.
The vacuum is a stationary point of the effective action, obtained
by integrating over the fermions only.
\begin{equation}
S_{\rm eff}=-\frac{1}{2G}\int d^4x\sigma^2-i{\rm Tr}\ln
\gamma_0\pmatrix{iV_d\cdot\partial &
      -\sigma(\vec q,x) \cr
-\sigma^{\dagger}(\vec q,x)& iV_u\cdot\partial\cr}.
\end{equation}
At the stationary points, the auxiliary field is given as
$\sigma_0(\vec q,x)=\left<\psi(\vec v_F^u,x)\psi(\vec v_F^d,x)\right>
\equiv\Delta(\vec q)$.
Since we are looking for a plane-wave condensate with a wave vector
$2\vec q$, $\sigma_0$ is translationally invariant.
The free energy density for the LOFF phase is then
given in the Euclidean space as
\begin{equation}
V(\Delta)=+{1\over 2G}\Delta^2-{1\over2}\int {d^4l\over (2\pi)^4}
\ln\left[1+{\Delta^2\over (l_0-il_u)(l_0-il_d)}\right],
\end{equation}
where we introduced new variables $l_u\equiv\vec v_F^u\cdot\vec l$ and
$l_d\equiv \vec v_F^d\cdot\vec l$.
Minimizing the free energy, we get the LOFF gap equation;
\begin{equation}
0={\partial V\over \partial \Delta(\vec q)}=
{\Delta(\vec q)\over G}-\int{{\rm d}^4l\over (2\pi)^4}
{\Delta(\vec q)\over (l_0-i l_u)
(l_0-i l_d)+\Delta^2}.
\label{gap}
\end{equation}
The characteristic feature of the gap equation (\ref{gap}) for the
LOFF paring is that the quark propagator is a function of three
independent momenta, $l_0$, $\vec l\cdot \vec v_F^u(\equiv l_u)$,
and $\vec l\cdot\vec v_F^d(\equiv l_d)$, while in BCS pairing it
is a function of two, $l_0$ and $\vec l\cdot\vec v_F$. In general,
we may decompose $\vec l$ as
$\vec l=\vec l_{\perp}+l_u\vec v_u^*+l_d\vec v_d^*$,
where $\vec v_{u,d}^*$ are dual to $\vec v_F^{u,d}$,
satisfying
$\vec v_F^a\cdot\vec v_b^*=\delta^a_b$ with $a,b=u,d$.
Though the magnitude of the Fermi velocity is $\vec v_F^a=1$ for
massless quarks, its dual has a magnitude $|\vec v_a^*|=(\sin\beta)^{-1}$,
where $\beta$ is the angle between $\vec v_F^d$ and $-\vec v_F^u$.

Since the quark propagator is independent of $\vec l_{\perp}$, it
just labels the degeneracy in the LOFF pairing. The perpendicular
momentum $\vec l_{\perp}$ forms a circle on the Fermi surface,
whose radius is given as $\mu_d\sin\alpha_d~(=\mu_u\sin\alpha_u)$,
where $\alpha_{d,u}$ are the angles between $\vec q$ and $\vec
v_F^{d,u}$, respectively. Upon integrating over $\vec l_{\perp}$,
the gap equation (\ref{gap}) becomes a (2+1) dimensional gap
equation. This is in sharp contrast with the gap equation in the
BCS pairing, which is (1+1) dimensional after integrating over the
$\vec l_{\perp}$, namely over the whole Fermi surface.

Integrating over $\vec l_{\perp}$, we find
the gap equation in Euclidean space to be
\begin{eqnarray}
1=
\int_{l_0,l_u,l_d}
{2G\mu_d\sin\alpha_d (3 \sin\beta)^{-1}\over (l_0-il_d)(l_0-il_u)+\Delta^2}
\simeq {2G\mu_d\sin\alpha_d\over3 \sin\beta}
\int_{\Delta}^{\Lambda^{\prime}}{{\rm d}l_0\over2\pi^3}
\left[{1\over2}\ln\left({l_0^2+\Lambda^2\over l_0^2}\right)\right]^2,
\end{eqnarray}
where $(\sin\beta)^{-1}$ arises from the Jacobian and we
introduced $\Lambda$ as the cutoff for $l_u,l_d$ and
$\Lambda^{\prime}$ for $l_0$. In the high density effective
theory, the expansion parameter is $|l^{\mu}|/\mu$. From the
condition that $|l^{\mu}|<l^2/\bar\mu$, where
$2\bar\mu^2=\mu_u^2+\mu_d^2$, we find the ultraviolet cutoff for
$l_{u,d}$ is $\Lambda=\bar\mu\sin^2\beta$. We also take
$\Lambda^{\prime}=\Lambda$, since the main contribution to the gap
comes from nearly on-shell quarks. Finally, integrating over
$l_0$, the gap equation becomes
\begin{equation}
1-{G_c\over G}={1\over 2}
\left({\Delta\over\bar\mu\sin^2\beta}\right)\,
\left[\ln\left({\Delta\over\bar\mu\sin^2\beta}\right)\right]^2,
\label{lgap}
\end{equation}
where the critical coupling for the LOFF gap
$G_c=3\pi^3\sin\beta/(4\mu_d\bar\mu\sin\alpha_d)$.
For a given $\delta\mu$ and $G$, the LOFF gap exists only when
$G>G_c$.

The best $q$, or equivalently the critical coupling for the LOFF pairing,
is determined
by minimizing the Free energy, obtained by integrating the gap
equation (\ref{lgap}) after doing the momentum integration,
\begin{eqnarray}
V(\Delta)=\int_0^{\Delta}{\partial V\over \partial \Delta^{\prime}}
d\Delta^{\prime}\simeq{1\over6G}\Delta^2\left(1-{G\over G_c}\right).
\end{eqnarray}
For $\bar\mu=400~{\rm MeV}$, $\delta\mu=30~{\rm MeV}$,
$\Delta_0=40~{\rm MeV}$, we have $\cos\beta=0.82$,
from which we obtain
$\Delta=0.076\bar\mu$ and $V=-1.9\times 10^{-5}~{{\bar\mu}^2/G}$.

\section{Asymptotic Density}

For the one-gluon exchange case, the calculation goes in parallel.
The Schwinger-Dyson (SD) equation for the quark two-point function is
given in the leading order in the hard-dense loop (HDL)
approximation as
\begin{equation}
\Delta({\vec q},l)=\left(-ig_s\right)^2
\int{{\rm d}^4k\over (2\pi)^4}V_u^{\mu}D_{\mu\nu}(l-k)V^{\nu}_d
{T^a\Delta({\vec q},k)T^a\over k\cdot V_dk\cdot V_u-\Delta^2({\vec q},k)},
\label{sd}
\end{equation}
where $T^a$ is the color generator in the fundamental representation
and $D_{\mu\nu}$ is the gluon propagator in the HDL approximation.
Since the Landau-damped magnetic gluons
give the dominant contribution, the SD equation becomes
in Euclidean space as
\begin{equation}
\Delta(\vec q)={2\over3}g_s^2\cot\beta
\int{{\rm d}l_0\over2\pi}{{\rm d}l_{\perp}\over2\pi}
{{\rm d}l_u{\rm d}l_d\over(2\pi)^2}
{1\over {\vec l}^2+{\pi\over4}M^2
|l_0|/|\vec l|}\cdot{\Delta(\vec q)
\over l\cdot V_E^u l\cdot V_E^d+\Delta^2},
\label{gap1}
\end{equation}
where the factor $2/3$ is due to the color factor
and $V_E^{\mu}\equiv(1,-i\vec v_F)$.
Since the high energy region
$\Lambda_0>|l_0|>2\Lambda_u$ does not contribute
the integration in Eq.~(\ref{gap1}) much,
the $l_0$ integration is regulated by the UV
cutoff of $l_u$ or $l_d$ integration. We then
note that when $l_u^2/(\sin\beta)^2>\alpha_s\bar\mu^2
|l_0|/|\vec l|$ the $l_u$ integration converges rapidly. Hence
for small coupling ($\alpha_s< 4 \sin\beta$)
we have a new UV cut-off for $l_u$ that satisfies
${\bar\Lambda_u}=\bar\mu\alpha_s\sin\beta$.
Integrating over $\vec l_{\perp},l_u$ and $l_d$, we get
\begin{eqnarray}
1={8\cot\beta\over 9\sqrt{3}\pi^2}\alpha_s^{2/3}{\bar\mu}^{-2/3}
\int_{\Delta}^{\bar\Lambda_u}{dl_0\over l_0^{1/3}}
\left[{1\over2}\ln\left(1+{\bar\Lambda_u^2\over l_0^2}\right)\right]^2
=\left({\alpha_s\over\alpha_c}\right)^{4/3}
\left[1-{2\over9}\left({\Delta\over\bar\Lambda_u}\right)^{2/3}
\left(\ln{\Delta\over\bar\Lambda_u}\right)^2
\right]
\label{gap2}
\end{eqnarray}
where
the critical coupling for the LOFF pairing by one-gluon
exchange interaction is
$\alpha_c=\left({\pi^2/2\sqrt{3}}\right)^{3/4}
\left({\sin\beta/\cos^3\beta}\right)^{1/4}$.

Now, let us calculate the vacuum energy to find the pair momentum
$2\vec q$ or the angle $\beta$ for a given parameters $\alpha_s$,
$\bar\mu$, and $\delta\mu$.
We find the vacuum energy is given as in the HDL approximation
$V(\Delta)\simeq-{\Delta^3/(6\pi^3)}
\left({\mu_dq/\bar\mu}\right)$.
The best $q$ or the angle $\beta$ is determined by minimizing the
vacuum energy for a given $\alpha_s$. For $\alpha_s=1$, we find
$\cos\beta=0.915$ and $V=-8.5\times10^{-4}\alpha_s {\bar\mu}^4$.
To find $\delta\mu_2$ for a given $\alpha_s$, we minimize the
vacuum energy with respect to $\beta$ and $\delta\mu$. At the
leading order the vacuum energy is minimized for large
$\delta\mu$. Therefore, to determine the correct $\delta\mu_2$ we
need to go beyond the current approximations and include the
$\delta\mu/\bar\mu$ corrections in the gap equation. This result
is consistent, at least qualitatively, with the observation in
\cite{Leibovich:2001xr} that the LOFF window widens considerably
in the one gluon exchange regime.

\section{Patching}

In the preceding analysis we assumed a simple plane wave
condensate $\Delta ( \vec{q} ) \propto e^{2 i q \cdot x}$.
Kinematical constraints require that quarks which participate in
this condensation lie on a ring-like region of volume $\sim
\Delta^2 \mu_i \sin \alpha_i$, where i = u,d.
However,
due to the rotaional invariance of the system any direction of
the pair momentum $2\vec q$ is possible. Thus, one needs
to patch condensates carrying same momentum but different
direction to increase the binding energy
The up and down quarks which pair in a LOFF condensate lie in
rings of radius $\mu_i \sin \alpha_i$, where $i = u,d$. These
rings lie in a plane perpendicular to $\vec q$, and have
thickness of order $\Delta$. If we sum over patches
associated with planar rotations of $\vec q$
without overlap, we will recover a fraction Fermi surface
which scales like $\mu^2$ and thus LOFF phase can be
compatible with BCS pairing.

\section{Conclusions}

We have used high density effective theory to study crystalline
superconductivity, using a local four fermion interaction to model
the interaction at intermediate density, and one gluon exchange
for the asymptotic regime. We obtained analytic results for the
condensate and vacuum energy, and conclude that the LOFF phase is
quite plausibly favored
$\Delta_0/\sqrt{2} ~<~ \delta \mu ~<~ \delta \mu_2$. We also
discussed how disjoint regions of condensate could be patched
together to obtain a lower vacuum energy. The optimal patching
configuration has yet to be determined for generic $\beta$, but it
seems possible that the binding energy of crystalline
superconductivity scales as the area of the Fermi surface ($\sim
\mu^2$) rather than as $\mu$ in the case of a plane wave LOFF
condensate.

\bigskip
I would like to thank F. Sannino and R. Ouyed for
the wonderful workshop.
I am grateful to M. alford, S. Hsu, K. Rajagopal, and Y. J. Sohn
for discussions and collaboration, upon
which this talk is based upon.
This work is supported by Korea Research
Foundation Grant (KRF-2000-015-DP0069).

\def\Discussion{
\setlength{\parskip}{0.3cm}\setlength{\parindent}{0.0cm}
     \bigskip\bigskip      {\Large {\bf Discussion}} \bigskip}
\def\speaker#1{{\bf #1:}\ }
\def\endDiscussion{}

\Discussion

\speaker{M. Mannarelli (INFN-Bari)}
You say that the LOFF phase appears for $g>g_c$
(in asymmetric quark matter). Is this the reason why in ordinary
BCS it has not been observed a LOFF phase?

\speaker{Hong}
For a given coupling $g$, there is always a window for LOFF as
$\delta\mu$
changes. However the window is larger for a stronger coupling.
It is therefore much easier to find a LOFF phase
in strongly coupled systems.

\speaker{M. Alford (Glasgow University)}
Just to clarify, for any coupling strength there is a range of $\delta\mu$
in which LOFF pairing is favored. This means that since $\delta\mu$ is
a function of $r$ in a neutron star, there is a reasonable chance of seeing
a shell of LOFF matter somewhere inside it.

\speaker{Hong}
For a given $\delta\mu$, the coupling between fermions has to be
larger than $g_c$ for a LOFF phase
to exist. This is the point I was trying to emphasize during my talk.
However, as the coupling becomes larger than
$g_{c1}~(>g_c)$, the BCS gap becomes large enough to win against a LOFF
phase.
A schematic phase diagram\footnote{
I thank M. Alford for helping me to understand the diagram in Fig.~1.}
as a function of the coupling $g$ and $\delta\mu$
can be given as Fig.~1:
\begin{figure}[htb]
\begin{center}
\epsfig{file=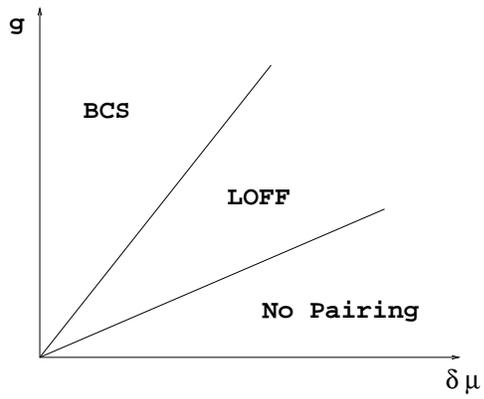,height=2in} \caption{A (schematic)
phase diagram in the coupling-$\delta\mu$ plane.} \label{fig:phase}
\end{center}
\end{figure}
\endDiscussion

\end{document}